**Częstochowski**

**Kalendarz Astronomiczny**

**2013**

**Rok IX**

**Redakcja**

**Bogdan Wszołek**
**Agnieszka Kuźmicz**

**Wersja elektroniczna kalendarza jest dostępna na stronach**

www.astronomianova.org
www.ptma.ajd.czest.pl

# Photometric Behavior of Five Long-Period Pulsating Stars

Larisa S. Kudashkina, Ivan L. Andronov, Lidia V.Grenishena

Odessa National Maritime University, Odessa, Ukraine
kuda2003@ukr.net, tt_ari@ukr.net

**Abstract**

The variation of average brightness during the time and the variation of the amplitude of the Mira-type stars T Cep, X CrB, U UMi, U Cyg, BG Cyg is studied. For the research, the observations of the members of the French association of observers of variable stars (AFOEV) covering almost 100 years are used.

All stars show cyclic variations of the specified parameters. For T Cep, U UMi and U Cyg, the values of quasi-period of the variations of average brightness are about 1400, 1025 and 1680 days, respectively. For the stars T Cep, U UMi and BG Cyg, the period of the brightness variations changes, but we used a mean one.

The dependences «average brightness - JD» and «amplitude - JD» are plotted. Results are discussed.

**Key words**

We have studied photometric behavior of five Mira-type stars (red giants of solar masses, AGB) on the basis of the own observations on negatives of the seven-camera astrograph of the Astronomical Observatory of the Odessa National University and on published observations of the French association of observers of variable stars (AFOEV).

For the time series analysis, we have used the software "Four" described by Andronov (1994) realizing trigonometric polynomial fit of statistically optimal degree with differential corrections to determine an optimal (for a given degree) value of the period. For determination of the dependence of the phase-averaged magnitude, we have used the method of "running sines" (see Chinarova 1998 for the first application). Reviews on used methods of time series analysis were presented by Andronov 2003, 2005). Recent our study of other stars was published by Kudashkina and Andronov (2010). The review on LPVs (Long-Period Variables) was presented e.g. by Kudashkina (2003).

For all stars, the basic photometric parameters have been calculated: the period $P$, amplitude, asymmetry of light curves. Also additional photometric parameters are calculated: amplitude of the first and second harmonics of mean light curve in stellar magnitudes ($r_1$, $r_2$), the phase of the maximum of the amplitude for each harmonic ($\Phi_{max}$) in respect to the maximum. The parameters of the slope of branches of the light curve are calculated: value of a slope of the ascending branch (incline) in the point of the greatest slope ($m_i$), similarly for a descending branch ($m_d$), phases of the greatest slope of branches, characteristic time of increase (decrease) of brightness by 1 magnitude ($t_i$, $t_d$), where $t_i=dt/dm$ at corresponding phases. Results are listed in Table 1.

Table 1. The characteristics of the mean light curve of the Mira-type stars.

| Star | T Cep | X CrB | U UMi | U Cyg | BG Cyg |
|---|---|---|---|---|---|
| $P$, d | 389.47±0.07 | 241.1±0.1 | 325.9±0.1 | 465.49±0.03 | 288.1±0.2 |
| $\Delta m$ | 1.72±0.04 | 4.33±0.02 | 2.63±0.02 | 2.95±0.02 | 2.3±0.1 |
| Asymmetry, $M-m$ | 0.47±0.01 | 0.506±0.006 | 0.504±0.005 | 0.47±0.01 | 0.42±0.03 |
| $m_{max}$ | 7.07±0.03 | 9.24±0.02 | 8.74±0.01 | 7.69±0.02 | 10.98±0.08 |
| $m_{min}$ | 8.79±0.04 | 13.6±0.3 | 11.38±0.02 | 10.65±0.03 | 13.3±0.1 |
| $JD_{max}$, 24….. | 0.263±0.003 | 49244.9±0.5 | 42769.4±1.3 | 45466.8±2.1 | |
| $m_i$ | 31.6±1.1 | 17.5±0.3 | 16.9±0.6 | 64.5±4.6 | 37.7±2.8 |
| $m_d$ | 39.3±1.9 | 16.9±0.3 | 27.4±1.6 | 60.8±4.5 | 56.1±9.3 |
| $t_i$ | 116±16 | 17.36±0.3 | 38.9±0.6 | 46±2 | 31±3 |
| $t_d$ | -83±8 | -18.03±0.3 | -39.7±0.6 | -96±7 | -60±12 |
| $r_1$ | 0.82±0.02 | 2.16±0.06 | 1.3±0.04 | 1.39±0.04 | 1.11±0.06 |
| $r_2$ | 0.06±0.02 | 0.13±0.01 | 0.11±0.01 | 0.12±0.01 | 0.17±0.06 |
| $\Phi_{max}$ ($r_1$) | 0.263±0.003 | 0.065±0.005 | 0.441±0.002 | -0.27±0.01 | -0.43±0.01 |
| $_{max}$ (r2) | 0.11±0.05 | 0.15±0.02 | 0.37±0.01 | -0.12±0.01 | -0.08±0.06 |
| $N_h/N_m$ | 0.60±0.28 | 0.37±0.24 | 0.42±0.25 | 0.23±0.48 | - |
| Type | I | I | I | II | - |
| $\rho_1$ | 0.40±0.15 | -0.1±0.23 | 0.98±0.15 | 0.84±0.13 | - |
| $\rho_1/\sigma_\rho$ | 2.71 | -0.4 | 6.53 | 6.35 | - |
| $\rho_2$ | -0.2±0.16 | 0.47±0.20 | -0.6±0.24 | -0.8±0.18 | - |
| $\rho_2/\sigma_\rho$ | -1.2 | 2.27 | -2.6 | -4.5 | - |

Parameters of the mean light curve for star BG Cyg have been determined on the observations on negatives of the seven-camera astrograph of the Astronomical Observatory of the Odessa National University. The interval of observations covers 11297 days (about 30 years). The phase light curves for all stars are shown on Fig. 1.

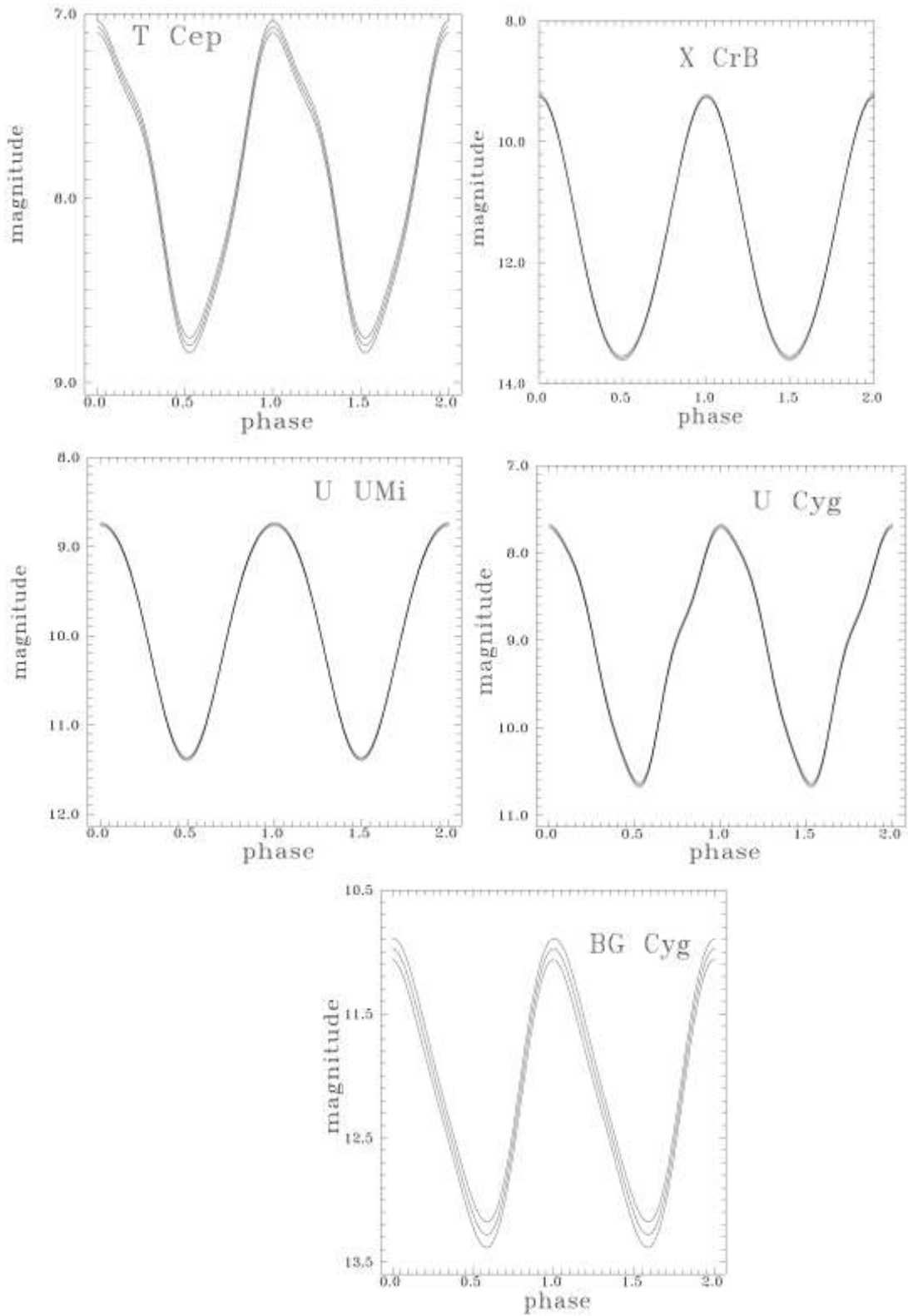

Fig.1. The approximations of the phase light curves using the statistically optimal trigonometric polynomial fit and corresponding ±1σ "error corridors". The parameters of the curves are given in Table 1.

The hump on an ascending branch is present at many long periodic variable stars. But it appears far not in each cycle. All investigated stars can be subdivided into two types: "Type I" contains stars which have a plenty of humps. Relative quantity of humps $N_h/N_m$ for them, as a rule, more than 0,3. "Type II"

213

contain stars with small quantity of humps or steps and for them $N_h/N_m \leq 0{,}3$. Here $N_h/N_m$ – the ratio of (observable) number of humps to (observable) number of maxima. Usually, the duration of a single hump is about 0.07 – 0.12 periods.

Also correlation coefficients between a phase of occurrence of a hump and duration of a hump ($\rho_1$) and between a phase of occurrence of a hump and average brightness of a hump ($\rho_2$) have been calculated. Possibly, presence or absence of correlation may be biased by selection effects (limiting amplitude of humps). However the tendency is observed that, for small amplitude stars, the duration of the hump is larger.

The time behavior of amplitude and average brightness was studied. The dependences are shown on Fig. 2-6.

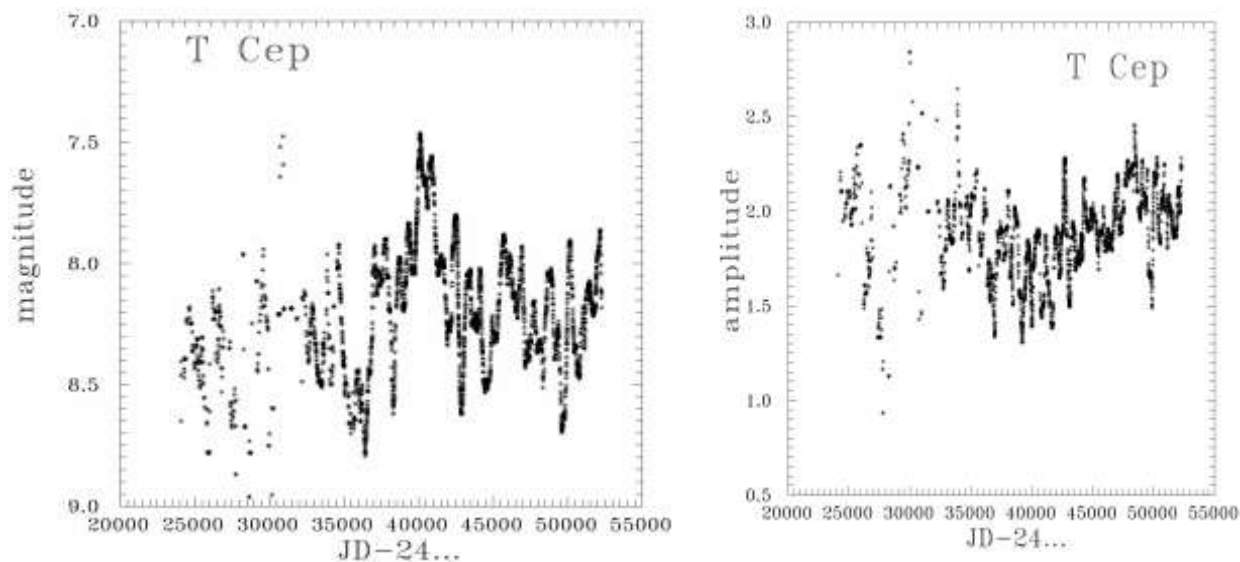

Fig.2. The average brightness changes with the cycle of about 1400 days.

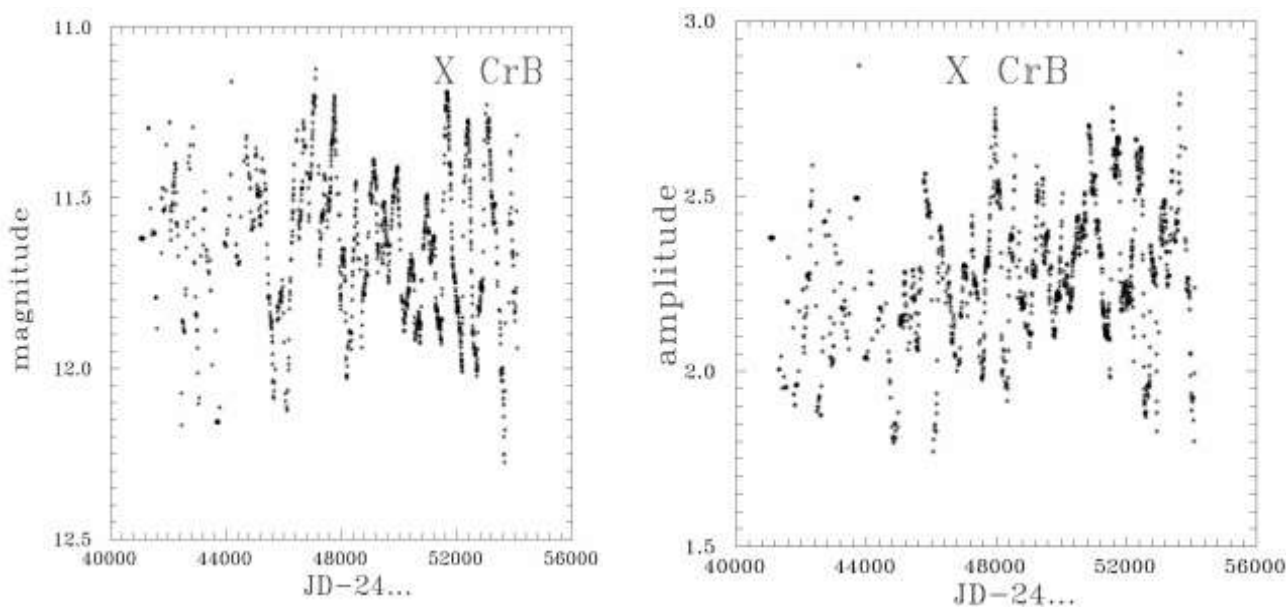

Fig.3. Value of a cycle of variations of average brightness and amplitude is not certain.

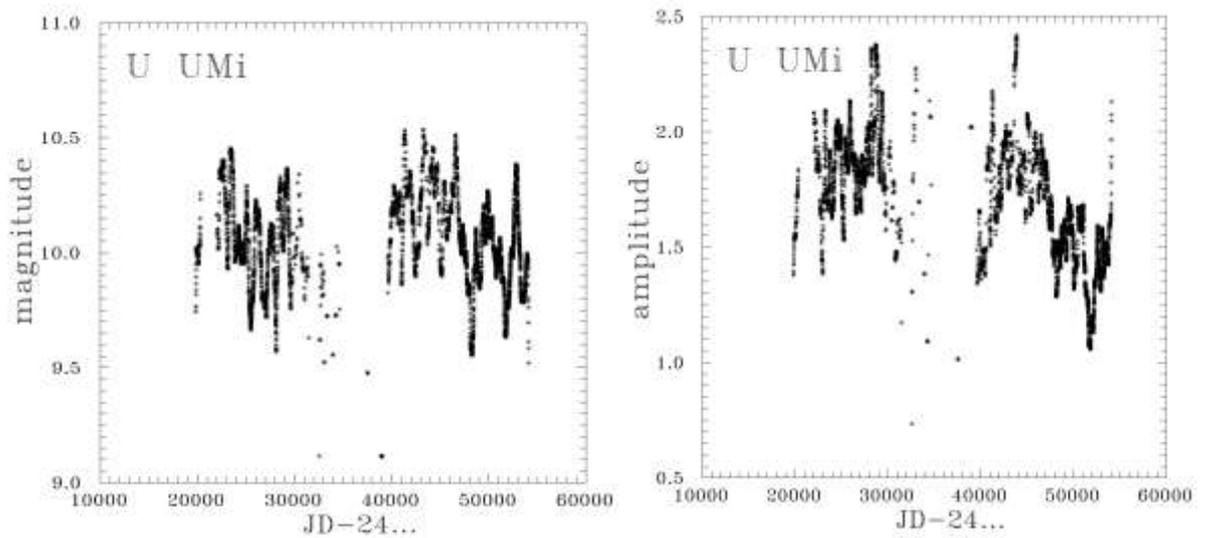

Fig. 4. The average brightness changes with the cycle of about 1025 days.

Most likely, variations of average brightness are caused by local processes in atmospheres and envelopes of stars and are not evolutionary. The same it is possible to tell about variations of amplitude. The shock wave from a radial stellar pulsation passes in layers with parameters changing by accidental image that is shown in chaotic or quasi-periodic variations of amplitude from a cycle to a cycle.

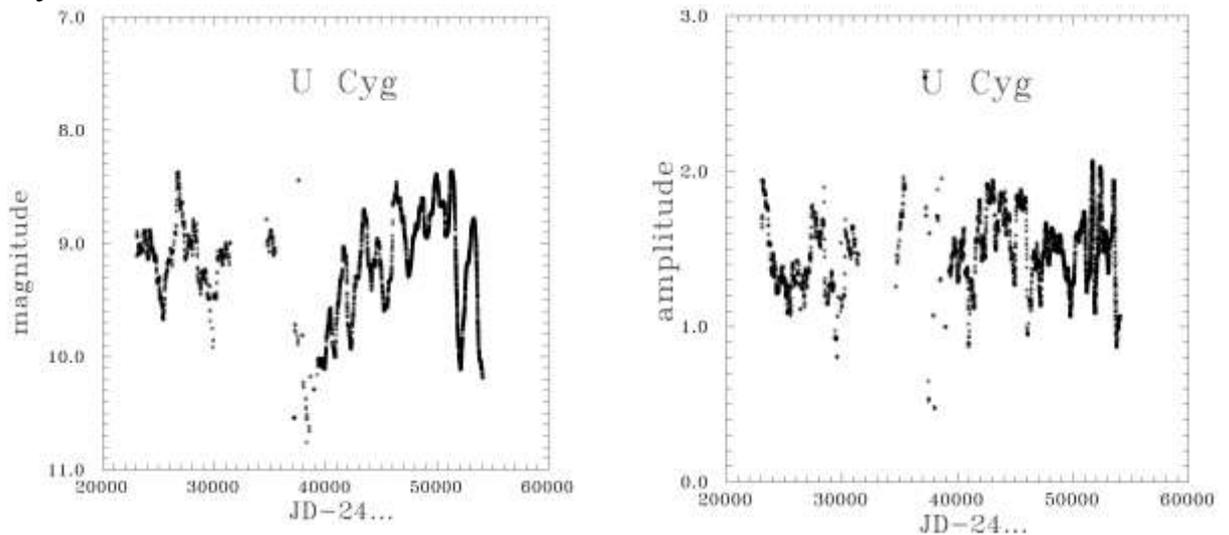

Fig.5. The average brightness changes with the cycle of about 1680 days.

215

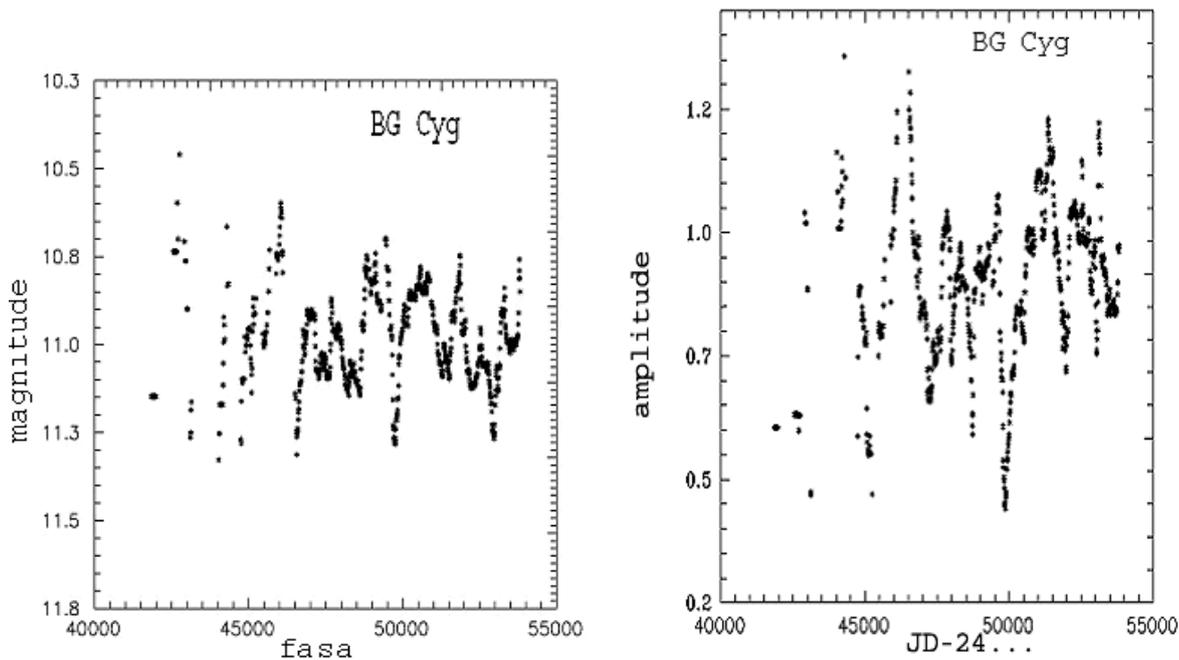

Fig.6. Value of a cycle of variations of average brightness and amplitude is not certain.

**Acknowledgements**

The study is based on the the AFOEF database (http://cdsarc.u-strasbg.fr/afoev/). This study is a part of the "Inter-Longitude Astronomy" (Andronov et al. 2010) and "Ukrainian Virtual Observatory" (Vavilova et al. 2012) projects. I.L.A. thanks Dr. Bogdan Wszołek for helpful discussions and hospitality during a stay in Częstochowa.

**References**

Andronov I. L., 1994, OAP, 7, 49A
Andronov I. L., 2003, ASPC, 292, 391A
Andronov I. L., 2005, ASPC, 335, 37A
Andronov I. L. et al., 2010, OAP, 23, 8A
Chinarova L. L., 1998, vsr. conf., 37C
Kudashkina L. S., 2003, KFNT, 19, 193K
Kudashkina L. S., Andronov I. L.; 2010, OAP, 23, 65K
Vavilova I. B. et al., 2012, KPCB, 28, 85V